\documentclass[12pt]{article}
\usepackage{amsfonts}
\usepackage{epsfig,amsmath}
\usepackage{setspace}
\usepackage{natbib}
\usepackage[usenames, dvipsnames]{color}
\numberwithin{equation}{section}
\usepackage{latexsym}
\usepackage{multirow}

\usepackage{chngcntr}

\newcommand{\bd}[1]{\ensuremath{\mbox{\boldmath $#1$}}}

\begin{document}
\doublespacing
\title{Bayesian adaptive distributed lag models}
\author{Alastair Rushworth}
\maketitle

\begin{abstract}
\noindent Distributed lag models (DLMs) express the cumulative and delayed dependence between pairs of time-indexed response and explanatory variables.  In practical application, users of DLMs examine the estimated influence of a series of lagged covariates to assess patterns of dependence.  Much recent methodological work has sought to develop flexible parameterisations for smoothing the associated lag parameters that avoid overfitting.  However, this paper finds that some widely-used DLMs introduce bias in the estimated lag influence, and are sensitive to the maximum lag which is typically chosen in advance of model fitting.  Simulations show that bias and misspecification are dramatically reduced by generalising the smoothing model to allow varying penalisation of the lag influence estimates.  The resulting model is shown to have substantially fewer effective parameters and lower bias, providing the user with confidence that the estimates are robust to prior model choice.  
\end{abstract}

\textbf{Keywords: Distributed lag model; penalized spline; adaptive smoothing}

\section{Introduction}
\label{sec:introduction}
Distributed lag models (DLMs) find application whenever the influence of a time-indexed independent variable is delayed and spread over time.  Although DLMs are widely applicable, their development was originally motivated by problems in econometrics (\citet{nerlove1958distributed}, \citet{almon1965distributed}, \citet{zellner1970analysis}, \citet{haugh1977identification}) and have more recently experienced a surge in popularity in environmental epidemiology (zanobetti, peng, welty, \citet{gasparrini2010distributed}).  Crucially, DLMs enable direct interpretation of the influence of a temporal exposure ensemble, which is particularly useful for characterising the total health impact of persistent environmental exposures such as air pollution or temperature (\citet{gasparrini2010distributed}, \citet{wyzga1978effect}).  Lag models can help to identify subtle types of time-dependence such as `mortality displacement', which occurs when exposure related mortality diminishes a vulnerable subpopulation, resulting in lower mortality in subsequent time intervals.  Mortality displacement has been widely documented (\citet{schwartz2000harvesting}, \citet{braga2001time}) and is conspicuous where lag influence is estimated to have a protective effect at low or moderate lags for what are otherwise harmful exposures (\citet{zanobetti2000generalized}).  Correctly identifying these effects depends on obtaining unbiased estimates of the underlying lag influence, and it is therefore essential that the DLM is correctly specified.  Since no previous study has formally quantified the estimation cost of model misspecification, this paper provides a comparison of various existing techniques.\\

If a time-varying explanatory variables $X_t$ has some influence on a response variable $Y_t$ over a set of lags $0,\ldots,p$, then a simple DLM can be expressed as the regression model $Y_t  =  \sum_{i=0}^{p} X_{t-i}\beta_i + \epsilon$, where $\epsilon$ is an independent error process and the parameters $\bd{\beta} = (\beta_0,\beta_1,\ldots,\beta_p)$ are referred to as the `lag curve' or lag function, as a function of $i$.  $\bd{\beta}$ cannot be reliably estimated using ordinary generalised linear models when $X_{t}$ and $X_{t-1}$ are correlated, and regularisation using shape constraints or smoothness restrictions over the lag parameters $\bd{\beta}$ are frequently used.  Recent studies have implemented various form of flexible parameterisation using polynomial splines and smoothness penalties, for example by combining uniformly spaced B-splines with ridge penalisation (\citet{zanobetti2000generalized}), logarithmically spaced B-splines (\citet{gasparrini2010distributed}), penalised B-splines (\citet{rushworth2013distributed}, \citet{obermeier2015flexible}), and Bayesian penalised splines with varying ridge penalties (\citet{muggeo2008modeling}).  Spline basis representations ensure a minimum level of smoothness, and offer computational relief when there are fewer knots than lags, which reduces the total number of parameters to estimate.  Although smoothness restrictions primarily serve to stabilise the model, they can also be exploited to incorporate additional structure over the shape of the estimated lag curve.  For example, it is widely assumed that the true lag curve has more curvature at short lags than at large lags, reflecting the intuition that recent events are generally the most influential, and influence should decay rapidly to zero for increasing lags.  Further structure can also be used to avoid discontinuity between the influence of the final modelled lag $\beta_p \neq 0$ and the implicitly assumed $\beta_{p+j} = 0$.  Some examples include \citet{koyck1954distributed} who restricted $\bd{\beta}$ to decay geometrically with increasing $p$; \citet{muggeo2008modeling} who combined a B-spline basis with parameters which were shrunk by a lag-varying ridge penalty; \citet{gasparrini2010distributed} used logarithmic knot placement in the lag dimension to enable varying smoothness; and \citet{welty2009bayesian} constructed a Gaussian process prior distribution that enforces smaller covariances on lag parameters at short lags than longer ones.  While each of these approaches achieves a level of lag-varying smoothness, each incorporates a parametric assumption on the type of penalisation that occurs, which may not be appropriate in general.\\

A further problem arises because the maximum lag $p$ must be chosen before model fitting: this is sometimes justified by drawing on previously published studies, for example, in several studies of the delayed effects of temperature on human health, $20\le p \le 30$ is consistently reported (\citet{braga2002effect}, \citet{armstrong2006models}, \citet{gasparrini2013reducing}).  In general, there is a cost to getting $p$ wrong: too large and the model risks overfitting, and too small can bias the resulting lag function estimate.  Few studies have investigated sensitivity of different model to choices of $p$, although several authors recognise that this is an ongoing difficulty for DLMs.   \citet{heaton2012flexible} and \citet{heaton2014extending} provide notable exceptions, and estimate $p$ as part of a two-stage approach.  However, DLMs are often applied in settings where the true lag curve is believed to decay gradually to 0, and in such settings, $p$ is likely to be poorly estimated and will lack a meaningful interpretation. \\

This paper proposes a pragmatic strategy that ensures flexible and automatic smoothing over the lag structure that avoids the need to either estimate or choose $p$ in advance.  By exploiting the property that very large lags have close to 0 influence, an automatic lag-dependent smoother is constructed that ensures strong penalisation at high lags so that $p$ can made arbitrarily large without the risk of overfitting.  This follows the tradition of richly specifying a hierarchical model and ensuring parsimony by liberal use of regularisation and smoothing.  A new DLM is described in Section \ref{sec:adaptivemodel} and in Section \ref{sec:simulations} a comprehensive simulation study illustrates the DLMs desirable properties, especially in comparison to existing approaches.  This study compares bias resulting from getting $p$ wrong, which is itself a central contribution, as DLMs have not previously been compared in this manner.  Section \ref{sec:discussion} concludes the paper with an overview of the main findings and a discussion of the implications for users of distributed lag models.

\section{Adaptive distributed lag models}
\label{sec:adaptivemodel}

Define $\mathbf{y} = (y_1,\ldots,y_n)$, a response variable time series of length $n$, and $\mathbf{x} = (x_1,\ldots,x_n)$, a time series of an exposure or input, whose influence on $\mathbf{y}$ up to some lag $p$ is to be estimated.  A Gaussian DLM for these data can be written as $y_i  \sim  \mbox{N}\left( \sum_{j=0}^p x_{i-j}\beta_j, ~\sigma^2\right)$ where $i= p+1,p+2, \ldots,n$, and $\sigma^2$ is the error variance.  The lag parameters $(\beta_0, \beta_1, \ldots,\beta_p)$ could be penalised and estimated directly, but for ease of comparison with existing approaches, these will be projected onto a B-spline basis of $K$ functions uniformly spaced over the range of $\mathbf{x}$.  The basis representation can be expressed as $\beta_j = \sum_{k=1}^K B_k(j)b_k$ where the further set of parameters $\mathbf{b} = (b_1,\ldots,b_K)$ determine the fitted lag curve.  The model can be expressed as $\mathbf{y}\sim \mbox{N}\left( \mathbf{X}\mathbf{b}, ~\sigma^2\right)$, where $\mathbf{X}_{ik} = \sum_{j=0}^p x_{i-j}B_k(j)$.  To control the roughness of $b_k$ in a way that enables variable smoothing, a similar approach to \citet{brewer2007variable} and \citet{reich2008modeling} is used, in which the intrinsic autoregressive (ICAR) prior distribution of \citet{besag1991bayesian} is generalised.   The conditional prior distribution for $\mathbf{b}$ under the ordinary ICAR is $\mathbf{b}\sim\mbox{N}(\mathbf{0}, \frac{1}{\lambda}\mathbf{P}^{-1})$ resulting in a conditional prior density of
\begin{eqnarray}
\pi(\mathbf{b}|\lambda) \propto \exp\left(-\frac{\lambda}{2}\mathbf{b}^{\top}\mathbf{P}\mathbf{b}\right) &=& \exp\left(-\frac{1}{2}\sum_{k=1}^{K-1} \lambda(b_{k+1} - b_{k})^2\right).
\label{eqn:icar}
\end{eqnarray}

The precision parameter $\lambda$ controls the extent to which the squared differences of consecutive pairs of $\mathbf{b}$ influences the posterior,  imparting a smoothing effect over the lag curve derived from $\mathbf{b}$.  Since each squared difference $(b_{k+1} - b_{k})^2$ contributes equally in the prior in in Equation \ref{eqn:icar}, variable smoothing can be achieved naturally by introducing a set of precision parameters $\bd{\lambda} = (\lambda_1,\ldots,\lambda_{K-1})$ that enables pairs of B-spline parameters to be smoothed more or less strongly.  This generalised prior can be written $\pi(\mathbf{b}|\lambda) \propto \exp\left(-\frac{1}{2}\sum_{k=1}^{K-1} \lambda_{k}(b_{k+1} - b_{k})^2\right) = \mbox{N}(\mathbf{0}, \mathbf{Q}^{-1})$ where $\mathbf{Q}$ is the precision matrix
\begin{equation*}
    \mathbf{Q} = \left(
      \begin{array}{cccccc}
        \lambda_1 & -\lambda_1 & 0 & 0 & \ldots & 0\\
        -\lambda_1 & \lambda_1 + \lambda_2 & -\lambda_2 & 0 &\ldots & 0\\
        0 & -\lambda_2  & \lambda_2 + \lambda_3 &  -\lambda_3 & \ldots & 0\\
        \vdots & \vdots &  \vdots & \vdots  &\ddots & \vdots\\
        0 & 0   & 0 &  0 & -\lambda_{K-1} & \lambda_{K-1} + \rho\\
      \end{array} \right).
  \end{equation*}

The additional parameter $\rho > 0$ added to the $(K,K)^{th}$ element of $\mathbf{Q}$ serves the dual purpose of ensuring $\mathbf{Q}$ is non-singular, and in providing an additional penalty that encourages the last element of $\mathbf{b}$ to approach zero.  This joint specification for $\mathbf{b}$ results in the conditional distributions
\begin{equation*}
    b_k~|~\mathbf{b}_{-k},\bd{\lambda}  \sim \left\{
      \begin{array}{ll}
      \mbox{N}\left(b_{k+1}, \frac{1}{\lambda_i}\right) & \textrm{if } k = 1\\
      \mbox{N}\left(\frac{\lambda_{k-1}b_{k-1} + \lambda_{i}b_{k+1}} {\lambda_k+ \lambda_{k+1}}, \frac{1}{\lambda_k + \lambda_{k+1}}\right) & \textrm{if } k = 2,\ldots,K-1\\
      \mbox{N}\left(b_{k-1}, \frac{1}{\lambda_{k-1}}\right) & \textrm{if } k = p.\\
    \end{array} \right.
  \end{equation*}

When $\lambda_1 = \lambda_2 = \ldots = \lambda_{K-1} = \lambda$ and $\rho = 0$, then $\mathbf{Q} = \lambda\mathbf{P}$, which shows that the ordinary ICAR prior in Equation~\ref{eqn:icar} is a special case of the adaptive prior.  A further smoothing prior distribution is placed over $\mathbf{\lambda}$ to restrict their flexibility - this is particularly attractive as it also reflects our intuition that the curvature of the lag function should not change too rapidly.  For this purpose, an ordinary ICAR prior is used to smooth $\mathbf{\lambda}$ such that $\bd{\tau}|\zeta^2 \sim \mbox{N}(\mathbf{0}, \zeta^2\mathbf{K}^{-1})$ where $\mathbf{K}$ is the $(K-1) \times (K-1)$ precision matrix
\begin{equation*}
    \mathbf{K} = \left(
      \begin{array}{rrrrrr}
        1 & -1 & 0 & 0 & \ldots & 0\\
        -1 & 2 & -1 & 0 &\ldots & 0\\
        0 & -1  & 2 &  -1 & \ldots & 0\\
        \vdots & \vdots &  \vdots & \vdots  &\ddots & \vdots\\
        0 & 0   & 0 &  0 & \ldots & 1\\
      \end{array} \right).
  \end{equation*}

Therefore, for small values of the variance $\zeta^2$, pairs of consecutive elements of $\bd{\lambda}$ are more strongly smoothed towards each other.  Finally, the variance parameters were assigned weakly informative priors $\sigma^2, \zeta^2 \sim  \mbox{Inverse-Gamma}\left(1, 1/2 \right)$.  The combination of adaptive smoothing with smoothed variances as described, results in a very flexible model that can accommodate complex shapes using a relatively simple specification.  The model described above assumes a Gaussian errors for simplicity, but the ideas easily generalise to accommodate other response distributions.\\

Inference for models fitted in the simulation study in Section \ref{sec:simulations} was achieved using Markov-chain Monte Carlo (MCMC) simulation using a mixture of Gibbs sampling and Metropolis-Hastings steps, using a mixture of optimised R (\citet{r}) and C++ functions implemented using the package Rcpp (\citet{eddelbuettel2013seamless}).  However, the models can be implemented using standard Bayesian model fitting software such as STAN (\citet{carpenter2016stan}).

\section{Simulation study}
\label{sec:simulations}

The adaptive model detailed in Section \ref{sec:adaptivemodel} was compared with several currently available distributed lag models, labelled $\mathcal{M}_1$, $\mathcal{M}_2$, $\mathcal{M}_3$, $\mathcal{M}_4$ and $\mathcal{M}_5$.  $\mathcal{M}_1$ is a non-adaptive version of the DL described in Section \ref{sec:adaptivemodel} (achieved by fixing $\lambda_1 = \lambda_2 = \ldots = \lambda_{K-1} = \lambda$), $\mathcal{M}_2$ is a Bayesian implementation of the DLM with linearly increasing ridge penalty proposed by \citet{muggeo2008modeling}, $\mathcal{M}_3$ is the new adaptive model described in Section \ref{sec:adaptivemodel}, $\mathcal{M}_4$ is the P-spline model of \citet{obermeier2015flexible} and $\mathcal{M}_5$ is the DLM with logarithmically spaced knots of \citet{gasparrini2010distributed}.   

\subsection{Comparison under different lag structures}
\label{sec:lagstructures}
\begin{figure}[h]
\centering
\includegraphics[width = 14cm]{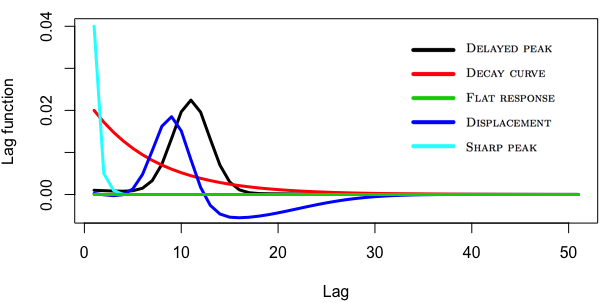}
\caption{The five lag curves used to simulate response and covariate data with different lag response.}
\label{fig:dlfunction}
\end{figure}

To compare estimation performance of models $\mathcal{M}_1 - \mathcal{M}_5$, time series response and covariate data were simulated assuming each of five different-shaped lag relationships, each with maximum lag $p=50$.  The curves are shown in Figure \ref{fig:dlfunction} and included a delayed peak scenario (\textsc{Delayed peak}), an exponential decay curve (\textsc{Decay curve}), a horizontal line at 0 equivalent to a null effect (\textsc{Null curve}), a curve exhibiting a shape similar to a mortality displacement effect (\textsc{Displacement}) and a sharply peaked function that declines rapidly to 0 (\textsc{Sharp peak}). \\

A covariate time series, $x_t$, was generated by assuming $x_t \sim \mbox{N}(0.5x_{t-1}, 0.1^2)$, and conditional on each lag function, a response series $y_t$ was generated using $y_t = \sum_{j=0}^{50} \beta_j x_{t - j} + \epsilon_t$ for $t = 51,\ldots,500$.  The error term $\epsilon_t$ assumed autocorrelation such that $\epsilon_t \sim \mbox{N}(0.2\epsilon_{t-1}, 0.1^2)$, whose parameters were selected for an acceptable signal-to-noise ratio.  The autocorrelation in both $x_t$ and $\epsilon_t$ reflects the temporal dependence in covariates and errors typically observed in real data.\\

Models $\mathcal{M}_1 - \mathcal{M}_5$ were fitted to each of 200 realisations of the processes $\{x_t,y_t\}$ described above. The models were assessed in terms of their ability to recover the true lag curve, as measured by root-mean squared error (RMSE) and squared bias (Bias$^2$) of the estimates, defined for the $j^{\textrm{th}}$ simulation as $\left(\frac{1}{51}\sum_{i = 0}^{50} (\hat{\beta}^{(j)}_{i} - \beta^{(j)}_{i})^2\right)^{1/2}$ and $\left(\frac{1}{512}\sum_{i = 0}^{51} \hat{\beta}^{(j)}_{i} - \beta^{(j)}_{i}\right)^{2}$, respectively.   Following \citet{hodges2001counting}, model complexity was recorded for fitted model $\mathcal{M}^{(j)}_i$ using the effective degrees of freedom $\textrm{ED} =  \textsf{tr}[\mathbf{X}(\mathbf{X}^{\top}\mathbf{X}+\hat{\mathbf{S}}_{\mathcal{M}_i^{(j)}})^{-1}\mathbf{X}^{\top}]$ where $\hat{\mathbf{S}}_{\mathcal{M}_i^{(j)}}$ is the estimated penalty component for the $i^{\textrm{th}}$ model fitted to the $j^{\textrm{th}}$ simulation.  In particular, $\hat{\mathbf{S}}_{\mathcal{M}_1^{(j)}} = \hat{\lambda}^{(j)}\mbox{diag}(1,2,\ldots,p+1)$; $\hat{\mathbf{S}}_{\mathcal{M}_2^{(j)}} = \hat{\lambda}^{(j)}\mathbf{P}$; $\hat{\mathbf{S}}_{\mathcal{M}_3^{(j)}} = \hat{\mathbf{Q}}^{(j)}$ and $\hat{\mathbf{S}}_{\mathcal{M}_4^{(j)}} = \hat{\lambda}^{(j)}\mathbf{P} + \hat{\rho}^{(j)}\mathbf{I}$.  For models $\mathcal{M}_1 - \mathcal{M}_4$ the model matrix $\mathbf{X}$ was identical, resulting from assuming $(2/3)\times (p+1) = 34$ uniformly spaced B-spline basis functions to represent the lag curve, such that the $(i, j)^{\textrm{th}}$ element, $\mathbf{X}_{i, k} = \sum_{j=0}^p x_{i-j}B_k(j)$.   Since lag curve smoothness for $\mathcal{M}_5$ is decided by finding the number of logarithmically spaced knots prior that minimises AIC, the effective dimension in this case is the number of knots, $\textrm{ED} =  n_z^{(j)}$ for the $j^{\textrm{th}}$ simulation.\\

Table \ref{table:sims} shows the average RMSE, Bias$^2$ and ED for each model and lag curve scenario combination.  The adaptive model, $\mathcal{M}_3$, outperformed all of the models in terms of RMSE across all lag curve scenarios.  Large separations were observed in model complexity across each scenario, with the adaptive smoothing $\mathcal{M}_3$ resulting in the lowest ED across all non-null scenarios, ranging between 3.86 to 7.97 compared to 6.4 and 19.09 for the other models.  A particularly concerning result, is that only $\mathcal{M}_3$ and $\mathcal{M}_4$ correctly result in degree of freedom $\approx 1$ under the \textsc{Flat response} scenario compared to a range of 7.7 and 18 for the others, which suggests that other approaches may falsely identify structure when none exists at all.  Bias was generally low across all models and scenarios, although higher values were noted across $\mathcal{M}_1$-$\mathcal{M}_4$ under the \textsc{Sharp peak} scenario.  As might be expected, the log-spaced knots of $\mathcal{M}_5$ performed favourably when lag influence declined rapidly from 0 (\textsc{Sharp peak}), and much less so when the strongest lag effect was larger than 0 (\textsc{Delayed peak}, \textsc{Displacement}).  The comparison clearly highlighted the need for care in choosing a lag smoother in practical application, and that the quality of fit strongly depends on the underlying lag function.  Furthermore, the simulations provide encouraging evidence that the newly proposed adaptive model is far more robust to these underlying differences, and might be a suitable choice when little is known about the structure being estimated.

\begin{table}[h!]
\centering
\begin{tabular}{llccccc}
\textbf{Lag curve shape} & \textbf{Summary} & $\mathcal{M}_1$ & $\mathcal{M}_2$ & $\mathcal{M}_3$ & $\mathcal{M}_4$ & $\mathcal{M}_5$\\
\hline
\multirow{ 3}{*}{\textsc{Delayed peak}} & RMSE & 17.31 & 14.53 & 10.32 & 15.51 & 22.68\\
& Bias$^2$ & 3.31 & 3.24 & 3.32 & 9.59 & 3.83\\
& ED & 18.53 & 14.99 & 7.12 & 12.14 & 15.81\\
 \hline
\multirow{ 3}{*}{\textsc{Decay curve}}& RMSE  & 16.57 & 11.59 & 8.45 & 9.46 & 14.15 \\
& Bias$^2$ &  1.45 & 1.65 & 2.32 & 4.30 & 0.79\\
& ED & 18.15 & 11.54 & 5.42 & 6.40 & 7.88\\
  \hline
\multirow{ 3}{*}{\textsc{Flat response}} & RMSE &  16.60 & 8.80 & 1.59 & 2.00 & 13.68 \\
& Bias$^2$ & 1.14 & 0.82 & 0.28 & 0.22 & 1.51\\
& ED & 18.07 & 8.08 & 0.77 &  0.68 &  7.74\\
  \hline
\multirow{ 3}{*}{\textsc{Displacement}} & RMSE &  17.32 & 14.13 & 11.24 & 15.29 & 20.34\\
& Bias$^2$ & 2.79 & 3.30 & 3.36 & 8.79 & 2.51\\
& ED & 18.55 & 13.97 & 7.97 & 12.78 & 13.37\\
  \hline
\multirow{ 3}{*}{\textsc{Sharp peak}} & RMSE  &  23.98 & 16.65 & 10.97 & 27.72 & 14.07\\
& Bias$^2$ & 16.01 & 11.63 & 6.81 &24.45 & 0.86\\
& ED & 19.09 & 11.05 & 3.86 & 10.85 & 7.94\\
  \hline
\end{tabular}
\label{table:sims}
\caption{Estimated lag function RMSE and effective degrees of freedom for four different types of lag curve shape, averaged across 200 simulated data sets.  RMSE and Bias have each been scaled by $10^3$.}
\end{table}

\subsection{DLM misspecification with maximum lag $p$}
\label{sec:misspec}
\begin{figure}
\centering
\includegraphics[width = 13cm]{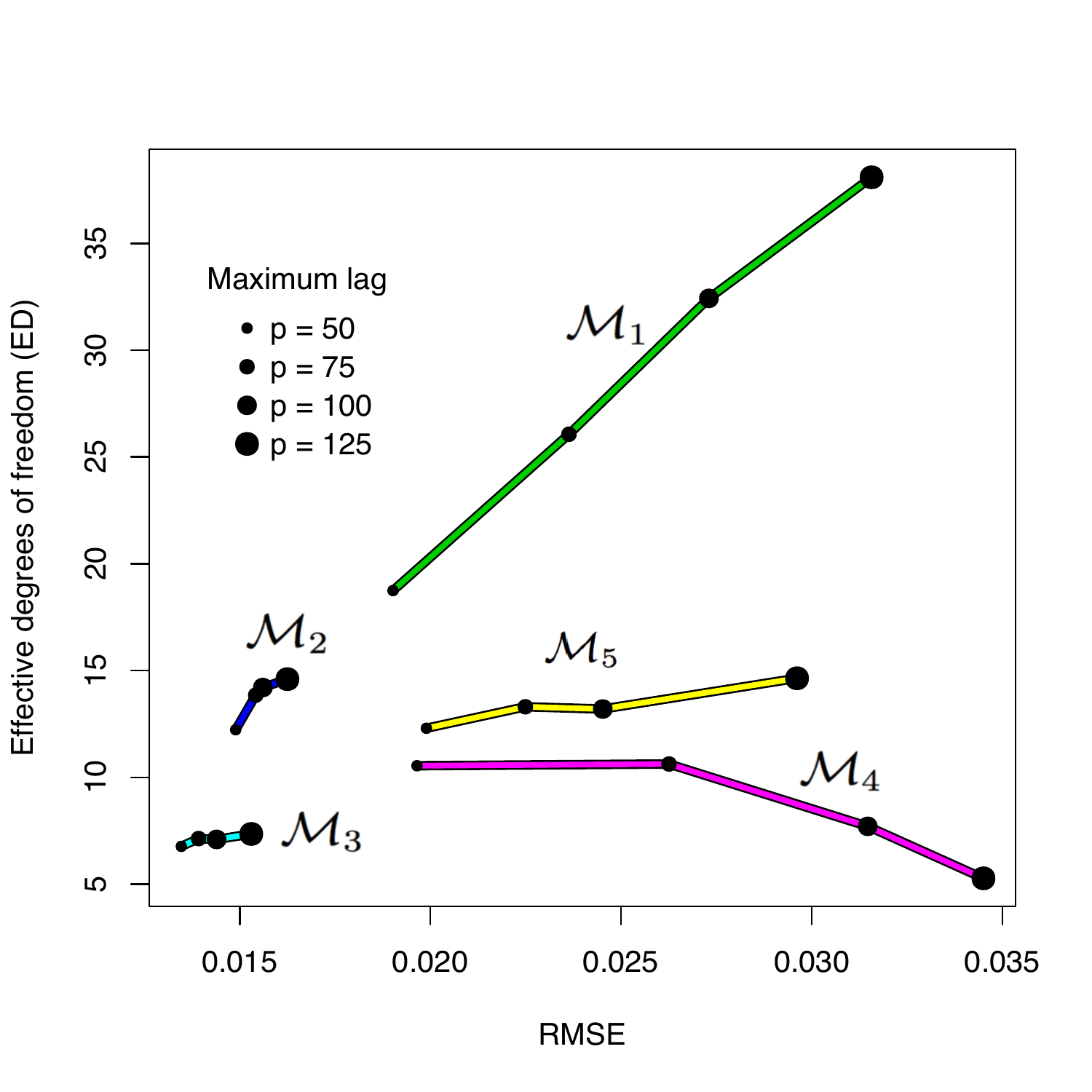}
\caption{Effective number of parameters and RMSE for each of $\mathcal{M}_1$ - $\mathcal{M}_5$.  Increasing diameter of the plotting symbols indicates larger numbers of maximum lags assumed in advance of model fitting.}
\label{fig:EDRMSE}
\end{figure}

Next, simulation was used to explore robustness of $\mathcal{M}_1$-$\mathcal{M}_5$ to different choices of maximum lag $p$.  The series $x_t$ and $y_t$ were generated exactly as in Section \ref{sec:lagstructures}, using only the \textsc{Displacement} lag function with 50 lags.  The models $\mathcal{M}_1$-$\mathcal{M}_5$ were fitted to the simulated $x_t$ and $y_t$, assuming maximum lags of $p = 50, 75, 100, 125$.  To ensure the fairest comparison, the range of flexibility for each model was restricted so that $0 < ED < 2p/3$, and also ensured that lag curve flexibility could grew proportionally with $p$.  For each of 200 fits, the RMSE of the lag curve estimate, ED and squared bias were recorded and the results are summarised in Figure \ref{fig:EDRMSE}.\\

Figure \ref{fig:EDRMSE} clearly illustrates, for $\mathcal{M}_1$ and $\mathcal{M}_5$ that ED increases steeply with $p$, even although the underlying true lag function is identical.  Furthermore, both the RMSE and ED are substantially smaller for the adaptive models $\mathcal{M}_3$ compared to any of the others in the comparison, and indicate much better overall performance.  The new adaptive approach $\mathcal{M}_3$ is more robust to user misspecification of `large' $p$, and the simulations raise concerns about the potential for overfitting using existing approaches.  However, it is noted that the effective degrees of freedom for $\mathcal{M}_3$ are not constant and do increase modestly with $p$, from 6.7 ($p=50$) and 7.4 ($p=125$).  Model $\mathcal{M}_2$ resulted in a comparatively favourable balance between RMSE and degrees of freedom, reinforcing the suggestion that adaptive models are preferable choices.

\section{Discussion}
\label{sec:discussion}

This paper makes several important contributions.  It was shown by simulation in Section \ref{sec:simulations} that estimation of lag structure can strongly depend on the type of smoothing model that is assumed, and that some existing approaches.  Several existing DLM models were shown to be non-robust to the choice of maximum lag $p$, even when the underlying lag function is identical, which suggests that the interpretation of lag estimates should be made with caution.  A new model has been proposed that combines automatic adaptive smoothing with a pragmatically large choice of $p$ to ensure simple and flexible smoothing of the lag curve that avoids sensitivity to the choice of $p$.  The new approach provides users of DLMs with a new way to explore their data with confidence that the estimates are not contaminated by artefacts that result from particular model choices. 
\clearpage
\bibliographystyle{chicago}
\bibliography{sdlm_nimrod}

\end{document}